\documentclass[preprint, aps,amsmath,amssymb,floats,mc
12pt]{revtex4}
\usepackage{amsmath}
\usepackage{graphicx}
\usepackage{dcolumn}
\usepackage{bm}
\usepackage[dvips]{color}
\begin{document}

\title{The Tsallis Parameter}
\author{ J. M. Conroy\thanks{Justin.Conroy@fredonia.edu} and H. G.
Miller\thanks{E-mail:hgmiller@localnet.com}}
\affiliation{Physics Department, SUNY
Fredonia, Fredonia, NY, USA}

\begin{abstract}
The exact solution of a particular form of the stationary state generalized
Fokker-Planck equations, which is given under certain conditions by the
classical Tsallis distribution, is compared with the solution of the
MAXENT equations obtained using the classical Tsallis entropy.  The solutions
only agree provided the Tsallis parameter, q, is no longer taken to be constant.

\end{abstract}

\maketitle

The nonextensive entropic measure proposed by Tsallis\cite{T88,IT89} introduces
a parameter, q, which is not defined\cite{T99,T09a}. It
has been argued that perhaps this parameter should be constant, if not
universally, at least for classes of dynamical systems\cite{GT04,WW00,TBT07}.
Attempts have been made to set limits on the value of this parameter \cite
{TBIB}. For the
practitioners this has been accepted de facto and calculations involving the
Tsallis entropy have generally used one piece of data  to determine the value of
q which
is then fixed over the range of interest\cite{GT04,TBIB}.  However, for the
generalized Fokker-Planck(GFP) equation\cite{PP95}, dissipative optical
lattices\cite{DBR06} and more recently in the
analysis of relativistic heavy ion collision \cite{WILK} q is most definitely
not a
constant, unless as  in the latter case, complicated scaling arguments are
invoked.

In the present letter we  consider  the GFP
equations \cite{R89} where an exact solution (both static and time dependent)
has been found and is shown under
certain circumstances to be equivalent to the Tsallis classical distribution
functions. In the static case we formulate and solve the equivalent MAXENT
problem\cite{PP99} for
different values of the equation of constraint. We solve for q in a manner
suggested earlier\cite{PMP04}  and find the numerical dependence of q on
the Lagrange multiplier associated with the equation of constraint.

Consider  the following non-linear one dimensional GFP equation\cite{PP95}
\begin{equation}
\frac{\partial F}{\partial t}=-\frac{\partial}{\partial
x}\{K(x)F\}+\frac{1}{2}Q\frac{\partial^{2}[F^{2-q}]}{\partial x^{2}},
\end{equation}
where F is the distribution function, Q is the diffusion coefficient, and K(x)
is the drift coefficient which determines the potential:
\begin{equation}
V(x)=-\int_{x_{0}}^{x} K(x)dx.
\end{equation}
The particular power $q-2$ is chosen in accordance with the quite general
discussion of the
generalized Bogulubov inequalities\cite{PT93} which points out that systems
which obey Tsallis statistics exhibit abrupt changes at $q=2$.

 Assuming the boundary conditions $F\rightarrow0$ for $x\rightarrow\pm \infty$,
the differential equation governing stationary solutions ($\frac{\partial
F}{\partial t}=0$) can be written as
\begin{equation}
K(x)F=\frac{1}{2}Q\frac{\partial (F^{2-q})}{\partial x}.
\label{GFP}
\end{equation}
By performing the redefinition $u=F^{2-q}$, separating variables and
integrating, one obtains the following stationary solution
\begin{equation}
F(x)=D[1-\beta(1-q)V(x)]^{\frac{1}{1-q}},
\label{Fq}
\end{equation}
where D is a positive (undetermined) integration constant.  Eq. (\ref{Fq}) is
precisely
the form of the distribution function obtained from Tsallis
Statistics\cite{T88,PP95}. One can
verify by substitution  that Eq.(\ref{Fq})  is a solution to Eq. (3) provided
\begin{equation}
\beta=\frac{2}{Q}[\frac{D^{q-1}}{2-q}]
\label{PLAST}
\end{equation}
Clearly, $\beta$ is a function of q.  It should be noted that this distribution
need not be normalized to 1.  Also, because the GFP Equation is nonlinear, for a
particular value of $\beta$ and q there is only one value of D that solves Eq.
(\ref{GFP}). Therefore, D is also an (undetermined) function of $\beta$ and q.
Note
that for $F(x)> 0$, $\beta> 0$ and q$< 2$. In the limit $q\rightarrow 1$ the GFP
reduces to the standard Fokker-Planck equation, $\beta\rightarrow \frac{2}{Q}$
and the stationary state adopts the usual form
\begin{equation}
 F(x)=D exp[-(2/Q) V(x)
\end{equation}

 The stationary state distribution can also be obtained from the classical
Tsallis entropy 
\begin{equation}
S_{q}=\frac{1}{q-1}(F-F^{q})
\end{equation}
via the  MAXENT equation\cite{PP99}
\begin{equation}
 \delta_F S_q =0
\end{equation}
along with the equation of constraint
\begin{equation}
Tr[F^{q}V]=C
\end{equation}
The solution of these equations yields the classical Tsallis distribution given
in Eq.(\ref{Fq}) for any allowed value of q.  In order to simultaneously
determine q an additional equation 
\begin{equation}
 \frac{\partial S}{\partial q}|_{\beta}=0
\label{dsdq}
\end{equation}
must be solved\cite{PMP04}. Note the distribution obtained is not normalized.

\begin{figure}         
\includegraphics[scale=.6,angle=0]{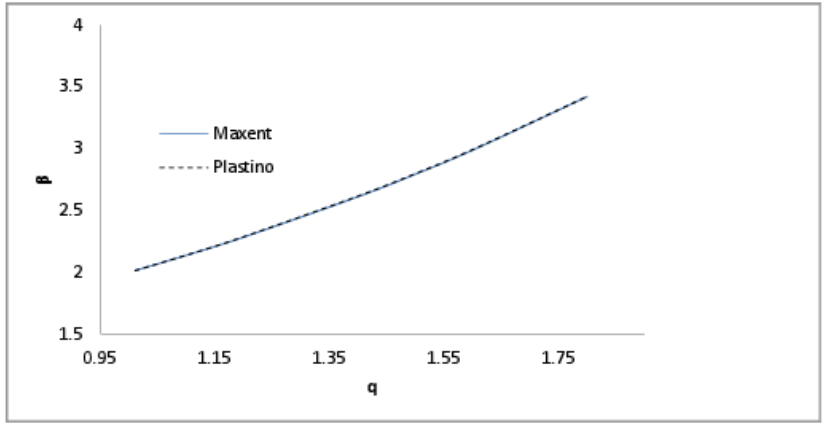}\center\vspace{-1mm}
             \caption{$\beta$ vs. q for V=1, and $\beta=\frac{2}{Q}D_{o}^{q-1}$ }
          \label{fig1}
       \end{figure}

For the simple case where V is a constant, the phase space
integrals in the trace each produce a constant factor which can be absorbed into
C, yielding
\begin{equation}
F^{q}V=C
\label{eqc}
\end{equation}
By solving Eq. (\ref{eqc}) for $\beta$ and substituting it into Eq. (8), we
numerically solve Eq. (\ref{dsdq}) for q.  FIG. (\ref{fig1}) is a plot of
$\beta$
vs. q for V = 1.

This curve can be fitted by the function $\beta=\frac{2}{Q}D_{o}^{q-1}$, where
$D_{o}=1.96$ and 	Q=1.  As already mentioned, D in Eq. (\ref{PLAST}) is an
undetermined function of $\beta$, and
therefore q.  To match the MaxEnt and GFP solutions, one must identify
$D=D_{o}(2-q)^{\frac{1}{q-1}}$.
\begin{figure}         
\includegraphics[scale=.9,angle=0]{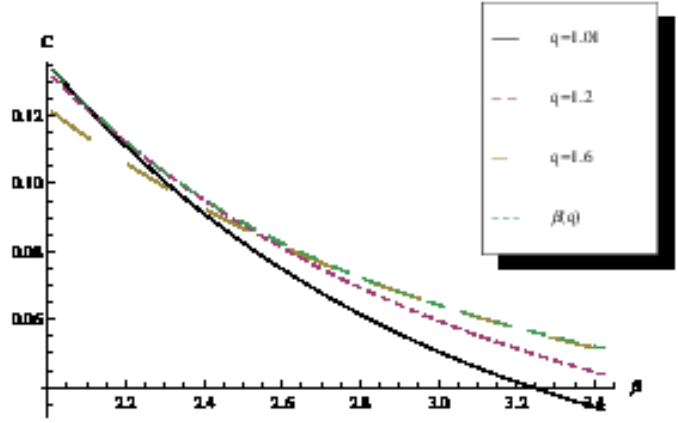}\center\vspace{-1mm}
             \caption{C vs. $\beta$ for V=1, $\beta(q)=\frac{2}{Q}D_{o}^{q-1}$ }
          \label{fig2}
       \end{figure}

Since q is a variable one cannot choose its value from one
data point and assume
it remains fixed over the entire range of interest.  To further illustrate this
point consider $C(\beta)$ as given by Eq. (\ref{eqc}), for constant V.  For a
thermodynamic system this can be thought of as the thermal response of the
system.  FIG. (\ref{fig2}) plots C vs. $\beta$ for  fixed values of q as well as
for $\beta=\frac{2}{Q}D_{o}^{q-1}$. Clearly the results obtained both in
magnitude and shape  do not agree well
except in the region around the  value of $\beta$ corresponding fixed
chosen value of
q.

Our results indicate that some care should be taken when the Maxent
equations are used with the Tsallis entropic measure  to describe experimental
data.  Although we have only considered the stationary solutions of a
particular form of the one dimensional GFP equation, the MAXENT solution is only
consistent with the exact solution when q is no longer treated as a constant.
Similar behavior is also to be expected in the solutions of the time dependent
GFP equation. Even if a scaling function can be found for every dynamical
situation,  the functional dependence of the unscaled Tsallis parameter should
be determined as a function of any Lagrange mutipliers associated with the
equations of constraint.
\vspace{5mm}

\noindent
Acknowledgment\\
We gratefully acknowledge some useful discussions with A. R. Plastino



\end{document}